\documentclass[a4paper,11pt]{article}

\usepackage{nfraconf} 
\usepackage{cite} 
\usepackage{graphics}

\begin{document}
\baselineskip=13pt

\title{Micro \& strong lensing with the Square Kilometer Array:\\ The
mass--function of compact objects in high--redshift galaxies}
\author{L.V.E. Koopmans} \address{Kapteyn Astronomical Institute\\
P.O.Box 800\\ NL-9700 AV Groningen\\ The Netherlands\\ E-mail:
leon@astro.rug.nl}

\author{A.G. de Bruyn} \address{Netherlands Foundation for Research in
Astronomy\\ P.O. Box 2\\ 7990 AA Dwingeloo\\ The Netherlands\\ E-mail:
ger@nfra.nl}

\maketitle \abstract{We present the results from recent VLA 8.5--GHz
and WSRT 1.4 and 4.9--GHz monitoring campaigns of the CLASS
gravitational lens B1600+434 and show how the observed variations
argue strongly in favor of microlensing by MACHOs in the halo of a
dark--matter dominated edge--on disk galaxy at z=0.4. The population
of flat--spectrum radio sources with micro--Jy flux--densities
detected with the {\sl Square--Kilometer--Array} is expected to have
dimensions of micro--arcsec. They will therefore vary rapidly as a
result of Galactic scintillation (diffractive and refractive).
However, when positioned behind distant galaxies they will also show
variations due to microlensing, even more strongly than in the case of
B1600+434. Relativistic or superluminal motion in these background
sources typically leads to temporal variations on time scales of days
to weeks. Scintillation and microlensing can be distinguished, and
separated, by their different characteristic time scales and the
frequency dependence of their modulations.  Monitoring studies with
{\sl Square--Kilometer--Array} at GHz frequencies will thus probe both
microscopic and macroscopic properties of dark matter and its
mass--function as a function of redshift, information very hard to
obtain by any other method.}

%%%%%%%%%%%%%%%%%%%%%%%%%%%%%%%%%%%%%%%%%%%%%%%%%%%

\section{Introduction}
 
Gravitational lens (GL) systems offer a versatile tool to study a
range of cosmological and astrophysical problems. One of most puzzling
and difficult problems to solve in all of astrophysics, is that of the
distribution and nature of dark--matter. Paczynski \cite{Pac86}
suggested to search for one of the dark--matter candidates --
i.e. massive compact objects, possibly stellar remnants or primordial
black holes -- using their gravitational lensing effect on background
stars. Many so called microlensing surveys have been undertaken since
then, with varying success \cite{Mao99}.

One major disadvantage of these Galactic microlensing surveys is the
low microlensing optical depth ($\tau$$\sim$$10^{-6}$)
\cite{Alcock97}. In multiply--imaged GL systems $\tau$$\sim$1, making
them much more suitable to search for microlensing variability
(e.g. \cite{Wambsganss90}).  The major disadvantage in this case is
the long variability time scale, which can amount to many years. This
is especially true if the lens galaxy is at a high redshift {\sl and}
the velocity of the microlensing magnification pattern is dominated by
the velocity of the compact objects with respect to the
line--of--sight to the stationary optical source.

What about microlensing in the radio? Most sources are much more
extended in the radio than in the optical. It is therefore often
believed that they are nearly unaffected by microlensing. However,
many flat--spectrum radio sources contain jet--components moving with
near or superluminal velocity.  These components can be as small as
several $\mu$as at the mJy level, to sub--$\mu$as at the $\mu$Jy
level. This is comparable or even smaller than the angular scale over
which the microlensing magnification pattern changes
significantly. Thus, at these levels radio microlensing can become
important.

In Sections 2--4, we will illustrate the above with the CLASS GL
B1600+434.  In Sect.5, we will show how the {\sl Square
Kilometer Array} can improve this in the future.

\begin{figure*}[t!]
\begin{center}
\resizebox{11cm}{!}{\includegraphics{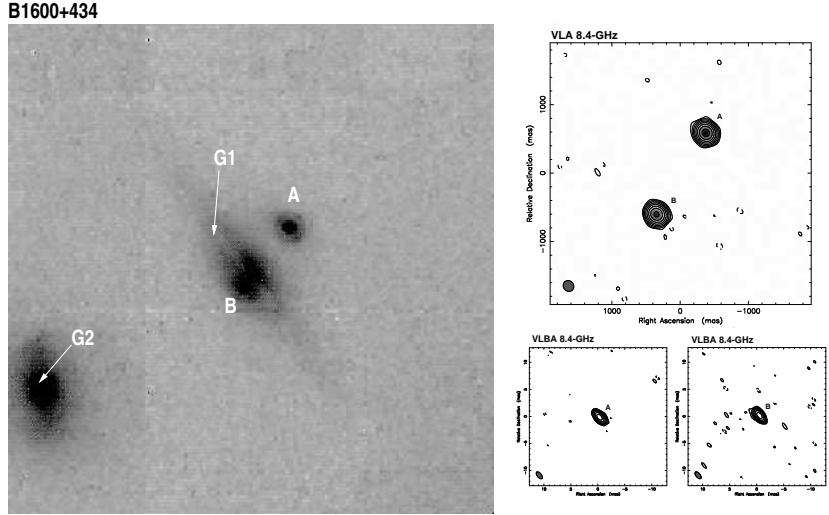}} 
\hfill
\parbox[b]{\hsize}{
\caption{\footnotesize {\bf Left:} HST NICMOS H--band image of
B1600+434, showing the two lens images (A and B), the edge--on spiral
lens galaxy (G1) and a companion galaxy (G2). Lens image B passes
predominantly through the disk/bulge of the spiral galaxy.  Image A,
however, passes exclusively through its dark--matter halo and is
therefore {\sl only} susceptible to microlensing by MACHOs, making
B1600+434 a unique target for this purpose.  {\bf Right:} Top: VLA
8.5--GHz A--array image of B1600+434.  Bottom: VLBA 8.5--GHz images of
the lens components B1600+434 A and B.  Both lens images show no sign
of any extended structure larger than $\sim$1 mas.}}
\end{center}
\end{figure*}

%%%%%%%%%%%%%%%%%%%%%%%%%%%%%%%%%%%%%%%%%%%%%%%%%%%

\section{The edge-on spiral lens system CLASS B1600+434}

The gravitational lens B1600+434 \cite{Jackson95} consists of a
1.4--arcsec double, lensed by a foreground edge--on spiral galaxy
(Fig.1; \cite{Jaunsen97}\cite{Koopmans98}), The lens galaxy and source
are at redshifts of 0.41 and 1.59, respectively
\cite{Fassnacht98}. B1600+434 was discovered in the {\sl Cosmic Lens
All--Sky Survey} (CLASS), whose mission is to find multiply imaged
flat--spectrum radio sources\cite{Browne98}.  Early VLA and WSRT radio
observations already indicated that the source was variable and
therefore suitable for determining a time delay \cite{Koopmans98}.  In
the coming paragraphs, we shortly describe constraints on the mass
model of the lens galaxy and the time delay between the lens
images. Subsequently, we show that most of the short--term variability
found in the radio light curves of the lens images is of external
origin.

{\bf \underline{Mass models}:} In \cite{Koopmans98} a range of mass
models to describe the edge--on disk galaxy (Fig.1) were presented,
with the aim of constraining the dark--matter halo around the lens
galaxy. A lower limit of $q_{\rm h}\ge$0.5 was found for the
oblateness of the dark--matter halo.  Moreover, at least half of the
mass inside the Einstein radius must be contributed by the
dark--matter halo, i.e.  the maximum--disk hypothesis does not apply
for this galaxy.  Assuming an isothermal halo model, a time delay of
about 54$h^{-1}_{50}$ days was predicted in a flat FRW universe with
$\Omega_{\rm m}=1$. The steeper MHP mass model for the dark matter
halo predict a time delay of about 70$h^{-1}_{50}$ days.  Once the
time delay between the lens images has been determined, constraints on
the Hubble parameter can be set.  However, more interesting, once the
Hubble parameter can be determined from several other JVAS and CLASS
lens delays, as well as other methods, the problem can be reversed and
constraints can be put on the mass profile of the dark--matter halo
around the edge--on spiral lens galaxy
\cite{Koopmans98}\cite{Koopmans99}.

{\bf \underline{Time-Delay}:} A monitoring campaign was initiated in
Febr. 1998 with the VLA in A-- and B--arrays at 8.5--GHZ
\cite{Koopmans99}.  B1600+434 was observed on average every 3.3
days. The radio maps have S/N--ratios of $\sim$300 and a resolution of
typically 0.2 (0.7) arcsec in A (B) array (Fig.1).  The final error on
the calibrated flux densities of the lens images is 0.7--0.8\%,
dominated by the uncertainties in the flux--density calibration. The
normalized calibrated light curves of both images are shown in
Fig.2. A 15--20\% long--term decrease in flux density of both images
is seen over the observing period of eight months, most likely
intrinsic source variability \cite{Koopmans99}.  One also sees short
term (days to weeks) variability of up to 10\% peak--to--peak in image
A, whereas the short term variability in image B is significantly less
(Fig.2).  We determined the time delay by scaling the observed
light--curve B by the instrinsic flux--density ratio (1.212) and
shifting it back in time, until the dispersion between the two image
light curves minimized. We find a time delay of $\Delta t_{\rm
B-A}$=$47^{+12}_{-9}$ days (95\% confidence; \cite{Koopmans99}). The
statistical error was determined from Monte--Carlo simulations and a
maximum systematic error of $-$8/+7 days was estimated.
Note that this method would have yielded a delay even if the source had not 
shown any short--term variability! 

{\bf \underline{External variability}:} Once the time delay and
flux--density ratio have been determined, the light curves can be
subtracted to see if any significant variability is left. In Fig.2 the
normalized difference light curve for B1600+434 is shown, using a time
delay of 47 days and a flux-density ratio of 1.212. The difference
light curve has an rms scatter of 2.8\%, which is inconsistent with a
flat difference light curve at the 14.6--$\sigma$ confidence level
\cite{KdeB99}. The individual normalized light curves (i.e. the light
curves divided by a linear fit) of images A and B (Fig.2) have rms
variabilities of 2.8\% and 1.6\%, respectively. Both are significantly
above the 0.7--0.8\%, expected on the basis of measurement errors.

\begin{figure*}[t!]
\begin{center}
\resizebox{10cm}{!}{\includegraphics{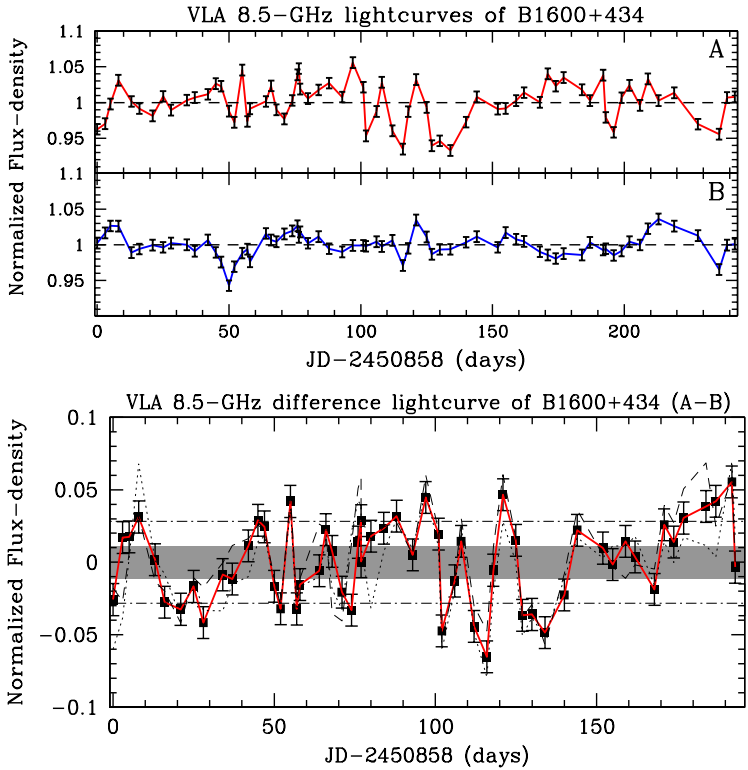}} 
\hfill
\parbox[b]{5.5cm}{
\caption{\footnotesize{\bf Top:} The normalized light curves of
B1600+434 A (upper) and B (lower), corrected for a long--term
gradient. The error on each light--curve epoch is 0.7 to 0.8\%.  {\bf
Bottom:} The normalized difference light curve between the two lens
images, corrected for both the time-delay (47d) and flux density ratio
(1.212).  The shaded region indicates the expected 1--$\sigma$ (1.1\%)
region if all variability were due to measurement errors. The
dash--doted lines indicate the observed modulation index of 2.8\%. The
dotted and dashed curves indicate the normalized difference curves for
a time delay of 41 and 52 days,
respectively.\medskip\medskip\bigskip}}
\end{center}
\end{figure*}

%%%%%%%%%%%%%%%%%%%%%%%%%%%%%%%%%%%%%%%%%%%%%%%%%%%

\section{Radio Microlensing by MACHOs in B1600+434?}

Below we investigate both scintillation and microlensing, as plausible
causes of the external variability in the lens images of B1600+434
\cite{KdeB99}.

\begin{figure*}[t!]
\begin{center}
\resizebox{\hsize}{!}{\includegraphics{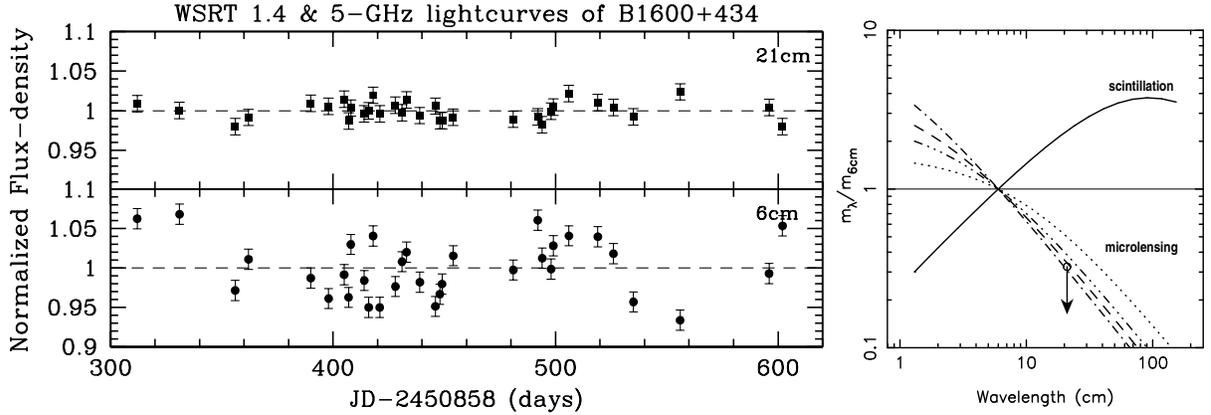}} 
\hfill
\parbox[b]{\hsize}{
\caption{\footnotesize{\bf Left:} The normalized WSRT 1.4 and 5--GHz
flux--density light curves of B1600+434. Only those epochs of the WSRT
light curves are shown that have both a 1.4 and 5--GHz measurement.
One notices a clear increase in the modulation index from 21 to 6 cm
by a factor $\sim$3.  {\bf Right:} Dependence of the modulation index
from scintillation and microlensing on wavelength. The solid line
shows the modulation index from scintillation. The broken lines show
the modulation indices from microlensing for several microlensing
models, constrained by the VLA 8.5--GHz light curves. All curves are
normalized to the modulation index at 6.0 cm observed with the WSRT in
1999. The open circle indicates the observed modulation index at 21
cm, agreeing remarkably well with that predicted from microlensing.}}
\end{center}
\end{figure*}

\underline{\bf Scintillation:} Scintillation, caused by the Galactic
ionized ISM, can explain both large amplitude variability in very
compact extra--galactic radio sources and lower amplitude `flicker' of
more extended sources (e.g. \cite{Rickett90}).  In B1600+434--A and B,
we observe several percent rms variability, possibly consistent with
`flickering'. The time scale is harder to quantify, but it appears
that we see variability time scales of several days up to several
weeks (Fig.2).

Let us summarize the arguments that argue against scintillation: (i)
The modulation index between the lens images differs, even though one
looks at the same background source. This requires significantly
different properties of the Galactic ionized ISM over scales of
1.4$''$.  (ii) Ongoing multi--frequency VLA and WSRT show a
decrease in the modulation index with wavelength (Fig.3), whereas the
opposite is expected for scintillation.  (iii) A modulation index of a
few percent corresponds to a variability time scale of less than 1--2
days for weak and strong refractive scattering, whereas variability
over much longer scales seems to be present.

Although the difference in modulation index can be explained by
considerable scatter--broadening of image B in the lens galaxy, all
evidence put together build a considerable case against scintillation.

\underline{\bf Microlensing:} There are several good reasons why
microlensing is a good explanation for the external variability that
is observed in B1600+434: (i) Because B1600+434 is multiply imaged,
the lens images pass through a foreground lens galaxy with
microlensing optical depths near unity. Combined with the fact that
many core--dominated flat spectrum radio source have superluminal
jet--structure (e.g. \cite{Vermeulen94}), makes it probable that
microlensing variability is occurring at some level on time scales of
weeks to months.  (ii) From microlensing simulations \cite{KdeB99}, we
find that a core plus a single superluminal knot can explain the rms
and time scale of variability in the lens images.  For image~A,
however, we require an average mass of compact objects
$\ge$0.5--M$_\odot$ to explain its significantly higher rms
variability. More complex jets, with multiple components, would of
course show reduced modulations, but this can then be compensated by
making them somewhat smaller (i.e. a larger Doppler boosting).  We
have tested this, by replacing the simple source with a real
jet--structure (i.e. 3C120), and shown that also more complex sources
give observable microlensing variations in their light curves.  (iii)
The apparent decrease in rms variability (i.e. modulation index) with
wavelength appears in agreement with a microlensed synchrotron
self--absorbed source that grows proportional with wavelength (Fig.3).

The current observations are all in agreement with the microlensing
hypothesis, even though some of the short--term variability maybe due
to scintillation \cite{KdeB99}. In fact, objects compact enough to be
microlensed should show scintillation at some level. In the case of
microlensing, caustic crossings can brighten objects orders of
magnitude more than scintillation if the emitting region is
significantly smaller than the Fresnel scale of several $\mu$as at
8.5--GHz. In this case we expect variability to be dominated by 
microlensing.

%%%%%%%%%%%%%%%%%%%%%%%%%%%%%%%%%%%%%%%%%%%%%%%%%%%

\section{B1600+434: Thus far...}

We have presented the first {\sl unambiguous} case of external
variability of an extra--galactic radio source, the CLASS
gravitational lens B1600+434 \cite{KdeB99}.  Several lines of evidence
indicate that it is not dominated by scintillation, but by
microlensing (in the lens galaxy) of a superluminal source.  This
requires a population of MACHOs in the halo around the edge-on disk
galaxy with masses $\ge$0.5--M$_\odot$.  The discovery of a population
of these objects might well indicate that its dark--matter halo mostly
consists of stellar remnants. Maybe these objects are the (white)
dwarfs as seen in the Hubble Deep Field \cite{Ibata99}
\cite{Mendez99}. Our ongoing multi--frequency WSRT (Fig.3) and VLA
observing campaigns should further tighten the constraints on this
tantalizing possibility.

%%%%%%%%%%%%%%%%%%%%%%%%%%%%%%%%%%%%%%%%%%%%%%%%%%%

\section{Future possibilities with the Square Kilometer Array}

Above we have shown an example of the current possibilities of
studying strong and microlensing in the radio. Preferably, one would
like to monitor these systems simultaneously at many different
frequencies and with much higher sensitivity. Although, we have
started a multi-frequency campaign of B1600+434 with the VLA, we are
working at the limits of present day technology.

The {\sl Square Kilometer Array} (SKA) has sensitivities of
sub-$\mu$Jy levels after a few minutes of integration, a resolution of
$\le$0.1$''$ at 20 cm and bandwidths around half a GHz (see
\cite{SKA}). With this sensitivity and the large instantaneous field
of view (FOV; around 1$^\circ \times$1$^\circ$ at 20 cm), SKA would
detect $\ge$100 radio sources per square arcmin (Fig.4; see also
Hopkins et al, this volume).

To estimate the number of GL system, useful for particular research
projects, we use the following formula:
\begin{equation}
    {\rm N}={\rm FOV} \times n_{\rm r} \times r_{\rm gl} \times f_{\rm
    r-type} \times f_{\rm g-type},
\end{equation} 
where FOV is the field--of--view, $n_{\rm r}$ is the number--density
of radio source on the sky, $r_{\rm gl}$ is the strong--lensing rate
of these sources, and $f_{\rm r-type}$ and $f_{\rm g-type}$ are the
fractions of radio--sources and lens--galaxies with properties that
one is interested in, respectively.  Using either the lensing rate in
the Hubble Deep Field, which has a comparable depth and
number--density of objects, or the typical lensing rate for high--z
sources of about $\le$$10^{-3}$, every synthesis observation with SKA
contains a few hundred to one thousand GL systems. About 10\% of these
radio sources are expected to have compact structure \cite{Hopkins99},
which will be easier to identify than the lensed starburst systems
which dominate the source counts at $\mu$Jy levels.  However, even
identifying the fainter cousins of the CLASS survey, will still be a
major task, with the typical distance between radio sources being
$\sim$6$''$, and not having the benefit of equally large optical
images to identify the lensing galaxies. However quad--configurations
and variability will be very helpful.

What to do with that many GL systems? Below we discuss two projects,
which we think might still be worth doing in $\sim$10 years time
(i.e. the development and implementation time--scale for SKA),
although there will undoubtedly be many more projects.

\begin{figure*}[t!]
\begin{center}
\resizebox{5.5cm}{!}{\includegraphics{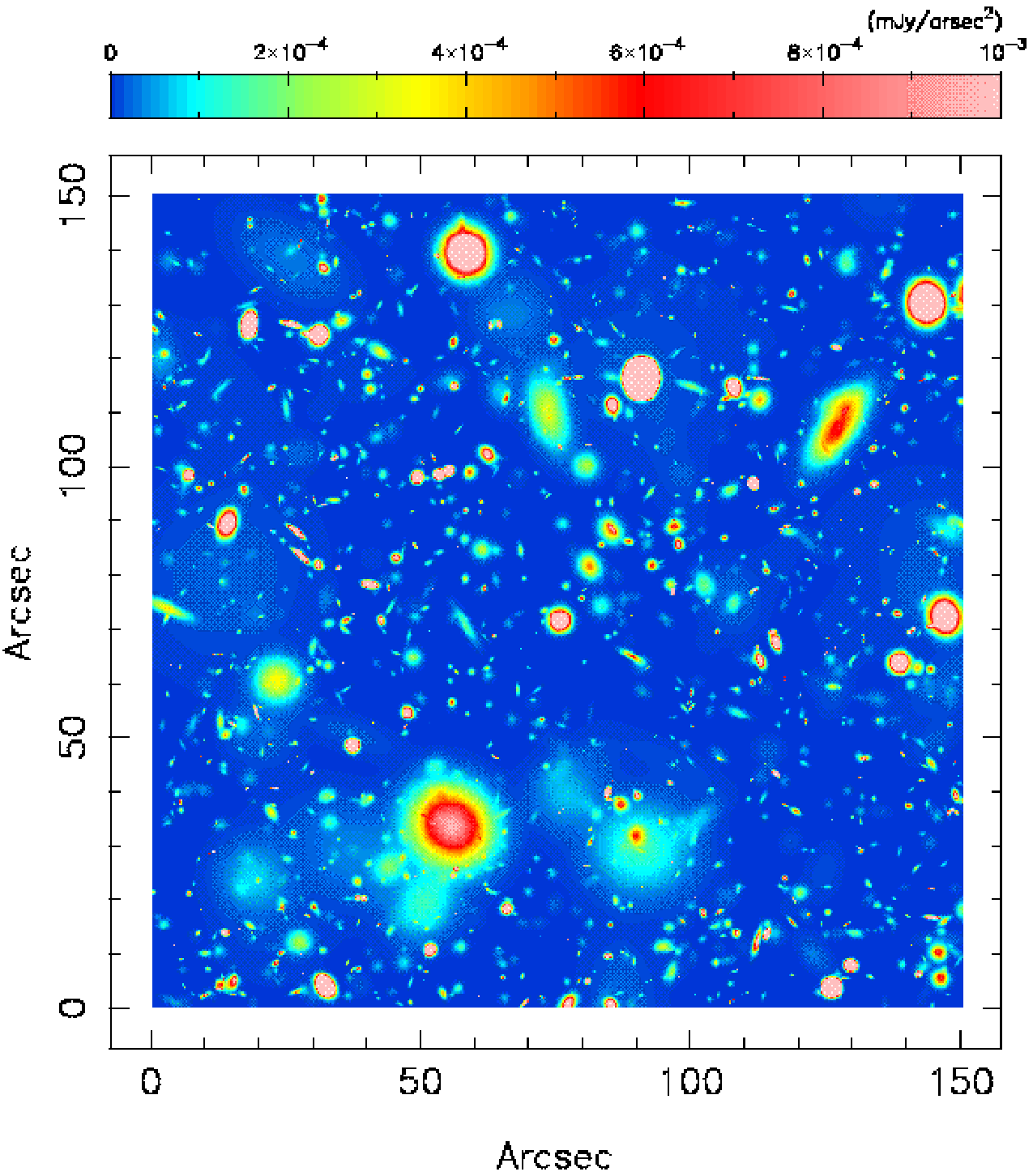}}
\resizebox{5.5cm}{!}{\includegraphics{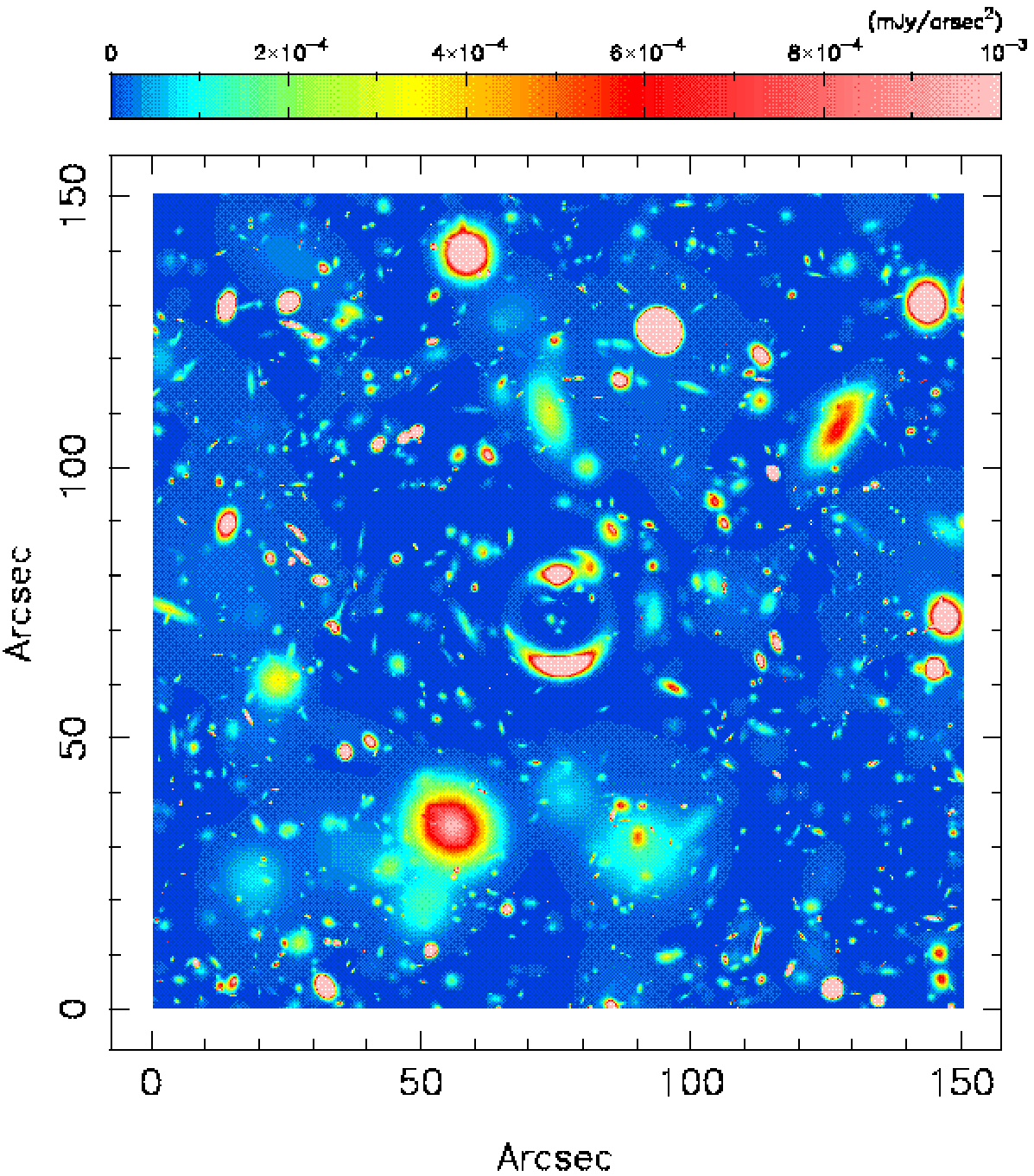}}
\hfill
\parbox[b]{4cm}{
\caption{\footnotesize Left: Simulation of a 150$''$$\times$150$''$ field,
as could be observed with the Square Kilometer Array in a single 12--h
integration at $\lambda$=20 cm (see also Hopkins et al, this
volume). The shown FOV is comparable to that of the Hubble Deep
Field. The real FOV of SKA at this wavelength will be around
1$^\circ$$\times$1$^\circ$. Right: The same, with a massive 
cluster (not shown) at $z$=0.5, lensing the background.\bigskip\smallskip} }
\end{center}
\end{figure*}

\underline{\bf Galaxy structure \& evolution at high redshifts:} To
study the structure \& evolution of galaxies up to very high
redshifts, one would preferably have both their colors {\sl and}
mass--distribution.  Colors can nowadays be obtain with instruments
such as HST and ground--based 8/10--m class telescopes. However,
obtaining information on the mass distribution of a significant sample
of high--redshift galaxies is and will remain exceedingly difficult in
the near future.  This is especially true for spiral galaxies. For
example, to obtain a decent sample of spiral galaxies with detailed
information on their mass distribution, one would like to have
Einstein--ring type GL systems (like B0218+357), which provide
considerable information on the mass distribution of the lens
galaxy. Taking a fraction of $\sim$15\% for the expected fraction of
spiral--lens galaxies, $\ge$50\% for the fraction of extended radio
sources (i.e. larger than the Einstein radius), $n_{\rm r}$$\approx$100
per sq. arcmin and $r_{\rm GL}$=10$^{-3}$, we expect $\sim$30 or more
spiral--lens galaxies with radio Einstein rings per FOV! Not only can
these Einstein--ring structures be used to constrain the mass
distributions of these spiral galaxies, they can also be used to probe
the HI content of the galaxy through absorption--line measurements,
simultaneously constraining their velocity fields.  Especially
HI--absorption against  lensed extended background radio sources would
be very informative (e.g. PKS1830-211; \cite{Chengalur99}).

\underline{\bf Constraining the mass-function of compact objects in
high--redshift galaxies:} As in the case of B1600+434, multi-frequency
radio monitoring of those lensed sources which are compact and
flat-spectrum, would allow one to separate variability resulting from
strong and weak scattering by the ISM, plasma lensing and
microlensing.  Time delays can be determined, allowing one to
determine the difference light curve between lens images.  This is the
only unambiguous method to separate intrinsic from non--intrinsic
variability. How important this will turn out to be remains to be
seen.

About 10\% percent of all disk--lens systems is expected to lie within
5 degrees of edge-on.  In fact, the CLASS/JVAS radio--GL survey has so
far discovered at least 17 new lens systems, of which B1600+434 is
one.  One can therefore expect $\sim$5\% of all GL to be of this
type. Assuming that $\sim$10\% of all radio sources are compact and
flat--spectrum, a few of these systems could be detected with SKA for
every FOV ($\sim$1 sq. degree).  That is, a survey with SKA could
detect $\sim$10 of these systems per day.  How could we use these
systems to learn about the mass function of compact objects in their
halo?

The study of variability in the lens images of these systems, as a
function of height above the disk of the lens galaxy, could provide
unique information on the composition of their dark--matter
halos. Although most lens systems would be expected to lie at
redshifts between 0.3 and 1, a fraction of $\sim$5\% will be found at
redshifts $\ge$1.5, allowing the study of the redshift dependence of
dark matter halos.

\begin{figure*}[t!]
\begin{center}
\resizebox{9cm}{!}{\includegraphics{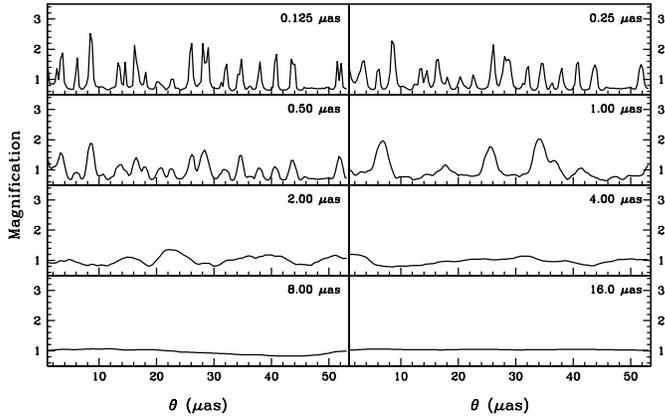}}
%\hfill
\parbox[b]{6.5cm}{
\caption{\footnotesize Arbitrary examples of simulated microlensing
light curves, showing the dependence on source size. Whereas in the
case of B1600+434, we are looking at jet--components of
$\ge$2\,$\mu$as, SKA will be looking at sources with
$\Delta\theta$$\ll$1$\mu$as.  The gain in picking up microlensing
induced variability is immediately
obvious. \bigskip}}
\end{center}
\end{figure*}

Because with SKA one is looking at sources at the $\mu$Jy level, the
expected angular size of these radio sources is
\begin{equation}
	\Delta\theta\approx\sqrt{\frac{S_{\mu\rm Jy}}{T_{12}}
	\left(\frac{\lambda} {25{\rm cm}}\right)^2}~~~\mu{\rm as},
\end{equation}
where $T_{12}$ is the brightness temperature of the source in units of
$10^{12}$\,K. If the source is has a redshift $z$ and
Doppler--boosting factor $\cal D$, one should replace $T_{12}$ with
${\cal D}\,T_{12}/(1+z)$.

Many compact flat--spectrum radio sources at high redshifts contain
superluminal components, with $\cal D$ of order a few. For example, a
superluminal 0.1\,$\mu$Jy jet--component with $\cal D$$\sim$10 in a
radio source at $z$=2 can be as small as 0.15\,$\mu$as (with
$T_{12}$$\approx$1), which is much smaller than the Einstein radius of
a 1--M$_\odot$ compact object at intermediate redshift!  We still do
not know whether the radio--faint AGN will also show superluminal
motion. On the other hand, much fainter galactic sources also show
superluminal motion.  Such a component moving with superluminal
velocity over a microlensing magnification pattern will show up in the
image lightcurves as variability of up to many tens of percents on time
scales of months, weeks and maybe even days. Because of the
compactness of these jet--components, the variability in their light
curves, induced by microlensing, will almost perfectly trace the
magnification pattern due to the compact objects, as is the case in
the optical (Fig.5).

Moreover, for superluminal sources the microlensing time--scale is not
dominated by the velocity of the compact lens objects (few
hundred km/s), but by the velocity of the jet--component
($\ge$$c$). Hence, not only the variability time--scale is compressed
by a factor $\ge$$10^{3}$, also the microlensing rate will enormously
increase \cite{Subramanian91}. Flat--spectrum radio sources could
therefore be the perfect probes to study compact objects and their
mass function in galaxies up to high redshifts!

However, how can one seperate microlensing, scintillation and
intrinsic variability of these sources? First, scintillation takes
place on time-scales of $\le$\,few hours for these ultra--compact
sources. By averaging over longer time scales one can easily remove
this, leaving only microlensing {\sl and} intrinsic variability.
Subtracting the averaged light curves of the different lens images, as
for B1600+434, will subsequently remove intrinsic variability. The
difference light curve that remains {\sl only} contains microlensing
variability.  Second, both microlensing and scintillation are strongly
dependend on frequency.  Whereas microlensing increases with
frequency, because the source becomes more compact (eqn.2),
scintillation will decrease, because of the frequency--dependent
refractive properties of the Galactic ionized ISM (e.g. Fig.3). The
scintillation is of course strongest around the transition frequency
from weak to strong, about 5 GHz. At longer wavelengths, e.g. 21cm,
very fast diffractive scintillation could also become important for
sub--microarcsecond sources. The large bandwith of SKA
($\Delta\nu$$\sim$$\nu/2$), however, will allow one to further
discriminate between microlensing and the different types of
scintillation.

In the case of B1600+434, separating properties of the source and
mass--function is complicated. In the case of very compact sources,
which will be observed with SKA, this is less problematic.  As
mentioned before, these sources are usually much smaller than the
angular scale over which the magnification pattern changes
significantly.  In other words, one can often regard the source as a
point source, which clearly simplifies unraveling the source and
mass--function properties. The statistical properties of the
difference light curves can thus be used to study the mass function of
compact objects at those positions where the lens images pass through
the lens galaxy (e.g. halo, disk or bulge).

Microlensing studies, of course, would not have to be confined to
multiply--imaged radio sources. Detailed multi--frequency studies of
suitable alignments of compact radio sources and foreground galaxies
could be used to study the properties of lensing matter as a function
of distance to the galaxy out to any distance, far into its
dark--matter halo!  A statistical study of the frequency--dependent
modulation characteristics as a function of distance to the foreground
galaxy can also be used to separate galactic scintillation from
microlensing because the galactic ISM is not expected to change
rapidly on angular scales of a few arcseconds.

%%%%%%%%%%%%%%%%%%%%%%%%%%%%%%%%%%%%%%%%%%%%%%%%%%%

\section{Conclusions}

Illustrated by the GL system B1600+434, we have shown the current
limitations, but also the exciting possibilities of detecting compact
objects in the dark--matter halo of intermediate--redshift galaxies
and contraining their mass function through radio microlensing.

The {\sl Square Kilometer Array} will be two orders of magnitude more
sensitive than the current VLA, and, because of multi--beaming, an
additional order of magnitude.  Not only can SKA find $\sim$10 GL
systems comparable to B1600+434 per day using its higher resolution,
larger bandwidth and continuous monitoring capabilities (because of
the multi--beam setup), we have also shown that microlensing,
scintillation and intrinsic variability can easily be separated in
these systems. SKA will therefore be a unique instrument to study
compact objects and their mass--function in galaxies and their
dark--matter halo up to very high redshifts. The large increase in the
microlensing rate (a factor $\ge$1000 compared with stationary optical
sources) for lensed flat--spectrum radio sources with superluminal
components makes this a nearly impossible task for any optical
telescope.

%%%%%%%%%%%%%%%%%%%%%%%%%%%%%%%%%%%%%%%%%%%%%%%%%%%

\end{document}